\documentclass{IEEEtran4PSCC}
\ifCLASSINFOpdf
   \usepackage[pdftex]{graphicx}
   \graphicspath{{./images/}}
\else
   \usepackage[dvips]{graphicx}
\fi
\usepackage[cmex10]{amsmath}
\makeatletter
\let\old@ps@headings\ps@headings
\let\old@ps@IEEEtitlepagestyle\ps@IEEEtitlepagestyle
\def\psccfooter#1{%
    \def\ps@headings{%
        \old@ps@headings%
        \def\@oddfoot{\strut\hfill#1\hfill\strut}%
        \def\@evenfoot{\strut\hfill#1\hfill\strut}%
    }%
    \def\ps@IEEEtitlepagestyle{%
        \old@ps@IEEEtitlepagestyle%
        \def\@oddfoot{\strut\hfill#1\hfill\strut}%
        \def\@evenfoot{\strut\hfill#1\hfill\strut}%
    }%
    \ps@headings%
}
\makeatother

\usepackage{amsmath}
\usepackage{amssymb}
\DeclareMathOperator{\E}{\mathbb{E}}
\usepackage{multirow}

\begin{document}
%
\title{Data-driven Linearization of Operational Dynamics of Soft Open Points in Power Distribution Networks}

\author{
\hspace{-1cm}\IEEEauthorblockN{Shoaib Hussain, Mostafa Farrokhabadi, Hamidreza Zareipour \\}
\vspace{-0.cm}
\IEEEauthorblockA{Department of Electrical and Software Engineering \\
The University of Calgary, Calgary, AB, Canada\\
}

}


\maketitle

\begin{abstract}
This paper proposes a novel data-driven linearization of operational power losses in a Soft Open Point (SOP). The proposed learning-based representation, combined with the standard segmentation of the quadratic device capacity inequalities, linearizes the operational dynamics of SOPs in unbalanced power distribution network applications without a significant drop in power loss estimation accuracy. The technique was tested in three applications using a three-phase extension of the IEEE 33-bus network. The case studies include a sizing application for a multi-terminal phase-changing soft open point (PCSOP), an operational optimization application for an active distribution network featuring a PCSOP, and a system simulation engine built for a Deep Reinforcement Learning (DRL) application that trains an agent to optimally dispatch a PCSOP in an unbalanced distribution network. The results of the three case studies were validated against the standard nonlinear AC load-flow representations and point to the efficacy of the proposed solution. 
\end{abstract}



\section{Introduction}
Power distribution networks are in the midst of an architectural transformation that is underpinned by a marked increase in the penetration of renewable energy sources (RES) and energy storage systems (ESS). The presence of intermittent power generators imposes several challenges when operating distribution networks. Distribution grids that have traditionally been designed for power flow from the generators to the load centers are today subjected to reverse flow of power caused by distributed energy resources (DERs) \cite{zhangDeepReinforcementLearning2021}. The fluctuations in power injection from RES often manifest as bus voltage-limit violations or congestion in the distribution network \cite{liDeepReinforcementLearningBased2022,sunTwoStageVoltVar2021}. Volt-VAR control and optimization is indispensable to system operation. Volt-VAR control is a mission-critical application that aims to deploy voltage and reactive power regulating devices such that the system keeps running without operational constraint violations \cite{gholamiStateoftheartTechnologiesVoltvar2022,liDistributedVoltVAROptimization2021}. Distribution system operators (DSOs) have had a number of tools at their disposal to control the operation of networks. This included assets such as On-load Tap-changing Transformers (OLTCs), Phase Shifting Transformers (PSTs), tap-changing series, and shunt compensators and Switchable Capacitor Banks (SCBs) \cite{yangOptimalReactivePower2017}. These devices were electromechanical in nature and operated on a slower time-scale. They were also subject to operational wear-and-tear that required periodic maintenance \cite{bazrafshanOptimalPowerFlow2019}. Amidst the widespread adoption of renewable distributed energy resources, DSOs are increasingly looking towards employing distribution-scale power-electronics-based control devices to mitigate operational challenges and exert effective control over the network \cite{urquizoUsingDFACTSMicrogrids2017,wangSummaryApplicationsDFACTS2012}. The distribution level variants of Flexible AC Transmission System (FACTS) devices, sometimes called D-FACTS, can be used to achieve several objectives. D-FACTS in distribution networks are used to compensate for phase imbalances \cite{achaFACTSModellingSimulation2010}, provide voltage and reactive power support \cite{botUsingDUPFCVoltage2015}, and facilitate the interconnection of distributed generators (DGs) and ESS \cite{changThreePhaseUnbalanceImprovement2022}. The integration of D-FACTS, however, adds another layer of complexity to the operation of distribution networks as the control variables of these devices need to be carefully tuned.

Soft Open Points (SOPs) are one of the newer members of power-electronic control devices that are commonly implemented using back-to-back voltage source converters (B2B-VSCs) \cite{fuadSoftOpenPoint2020}. The adoption of SOPs in distribution networks worldwide has steadily grown since their inception in 2001 \cite{jiangOverviewSoftOpen2022}. SOPs are typically employed as static replacements of Normally Open Points (NOPs) in radial distribution networks and have a much faster response time compared to traditional mechanical switches. SOPs provide control over active and reactive power injections when used to interconnect two or more feeders of a distribution network \cite{louNewPhaseChanging2020}. SOPs can be classified based on either the number of feeder terminals connected or the number of phases involved. The simplest implementation of an SOP is a single-phase two-terminal device whereas the three-phase phase-changing SOP (PCSOP) is the most sophisticated one to date \cite{louNewPhaseChanging2020}. SOPs function by transferring AC energy from network laterals to a common DC link and then back to AC again using VSCs. PCSOPs additionally have the capacity to transfer power from one phase of a feeder to either the same or a different phase of another feeder. The active and reactive power injection variables are interrelated due to the capacity constraints of the B2B-VSCs. A balance must be maintained for the active power transferred across feeders in addition to the capacity constraints. This leads to firm equality (active power balance) and operational inequality (active-reactive capacity) constraints that must be maintained to avoid limit violations \cite{diaaeldinOptimalNetworkReconfiguration2019,caoOperatingPrincipleSoft2016}.

The SOP-based planning and operational optimization schemes proposed in the literature are presented on the basis of mathematical optimization. This includes mathematical programming techniques such as Mixed Integer Linear Programs (MILP), Mixed Integer Second Order Conic and Semi-definite Programs (MISCOP, SDP), Non-linear and Mixed Integer Non-linear Programs (MINLP), and heuristic optimization techniques. Mathematical optimization is the prevailing paradigm for implementing many distribution system management applications. Stochastic mathematical programming techniques are powerful tools for distribution expansion planning and sizing and siting applications involving SOPs, DGs, and ESS. A number of optimization techniques have been proposed for optimally dispatching SOPs in distribution networks. Some of the recent considerations include: a distribution planning model that optimizes the sizing and siting of ESS in networks incorporating SOPs \cite{baiDistributedEnergyStorage2018}; a siting and sizing MINLP application for SOPs and DGs in reconfigurable grids \cite{diaaeldinOptimalNetworkReconfiguration2019}, and an SDP relaxation-based operational optimization model designed to mitigate phase imbalances using PCSOPs \cite{louMultiterminalPhasechangingSoft2022}, among others.

\begin{figure}[!t]
  \centering
  \includegraphics[scale=0.375]{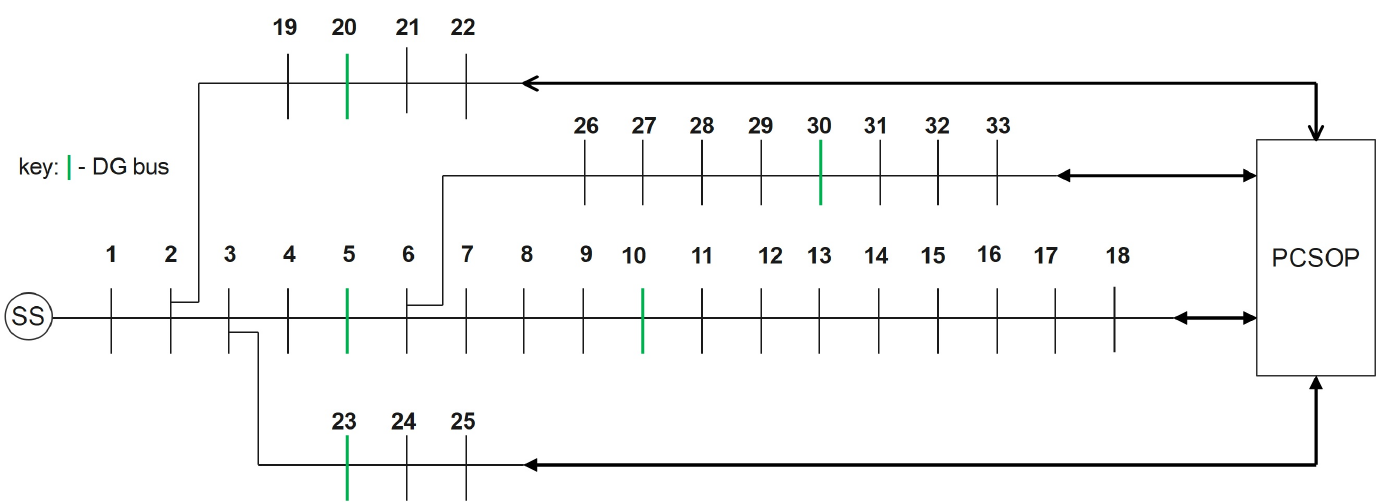}
  \caption{IEEE 33-bus test-feeder with DGs and PCSOP}
  \label{fig:fig1}
\end{figure}

SOPs are lossy devices where the per-phase loss is represented as a nonlinear function of active/reactive power injections. The intertwined capacity and device power loss expressions yield nonlinear programming formulations that can slow down or stall load-flow and optimization routines. The nonlinear representation of device dynamics limits the versatility of optimal planning and dispatch applications as well as data-driven control algorithms such as reinforcement learning that require repeated solves of the load-flow routines. In light of the drawbacks associated with the standard operational representation of SOPs, we propose a novel data-driven, regression-based reformulation of the SOP power-loss functions. The linearization technique transforms the nonlinear power-loss expressions into a linear combination of the active and reactive power injection variables. The proposed procedure, combined with the well-known segment-wise representation of quadratic device capacity constraints, linearizes the operational dynamics of SOPs without a significant loss in accuracy. The rest of the paper is organized as follows. Section II presents the proposed functional approximation of SOP power losses, the associated linearization procedure, and the mathematical modeling of the case-study applications that exemplify the proposed representations. Section III presents the numerical results of the case studies and section IV concludes the paper.        

\begin{figure}[!t]
  \centering
  \includegraphics[scale=0.5]{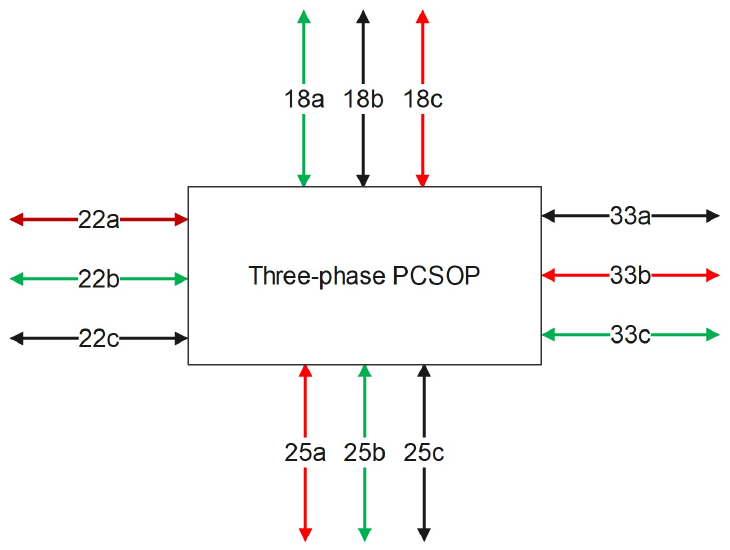}
  \caption{PCSOP with color-coded Tie-groups}
  \label{fig:fig2}
\end{figure}

\section{Methodology}
A three-phase extension of the IEEE 33-bus test feeder integrated with a PCSOP was used as a testbed for all case studies. The unbalanced distribution network also hosts several DGs as depicted in figure 1. Figure 2 shows the specifics of the three-phase bus connections to the PCSOP. The PCSOP in this setup is used to route active power across phases and feeder laterals. The color-coded tie-groups depicted in figure 2 indicate the phases involved when active power is transferred between buses of the network laterals. With reference to figures \ref{fig:fig1} and \ref{fig:fig2}, the operation of a SOP in a distribution network is governed by a set of equality (active power balance) and inequality constraints related to the functional capacity of the B2B-VSCs \cite{caoOperatingPrincipleSoft2016}. The active power and capacity constraints are given by expressions in (20) and (21), respectively.
\begin{equation}
\label{eq:1}
\begin{aligned}
Tie-group &\ 1: P^{22}_a+P^{18}_c+P^{33}_b+P^{25}_a\\
&+P^{22}_{loss,a}+P^{18}_{loss,c}+P^{33}_{loss,b}+P^{25}_{loss,a}=0 \\
Tie-group &\ 2: P^{22}_b+P^{18}_a+P^{33}_c+P^{25}_b\\
&+P^{22}_{loss,b}+P^{18}_{loss,a}+P^{33}_{loss,c}+P^{25}_{loss,b}=0 \\
Tie-group &\ 3: P^{22}_c+P^{18}_b+P^{33}_a+P^{25}_c\\
&+P^{22}_{loss,c}+P^{18}_{loss,b}+P^{33}_{loss,a}+P^{25}_{loss,c}=0
\end{aligned} 
\end{equation}

\begin{equation}
\begin{aligned} 
\label{eq:2}
(P^{\chi}_{\phi})^2+(Q^{\chi}_{\phi})^2 \leq (S^{\chi}_{\phi,max})^2 \\
\forall\phi \in (a,b,c),\ \chi \in (22,18,33,25) 
\end{aligned}
\end{equation}
The power loss terms in (\ref{eq:1}) are often expressed in terms of the active and reactive power injections and represent the leakage in power transferred across feeders \cite{jiangOverviewSoftOpen2022,louMultiterminalPhasechangingSoft2022}.
\begin{equation}
\begin{aligned}    
\label{eq:3}
P^{\chi}_{loss,\phi} = \Gamma\sqrt{(P^{\chi}_{\phi})^2+(Q^{\chi}_{\phi})^2}
\end{aligned}
\end{equation}

\begin{figure}[!t]
  \centering
  \includegraphics[scale=0.58]{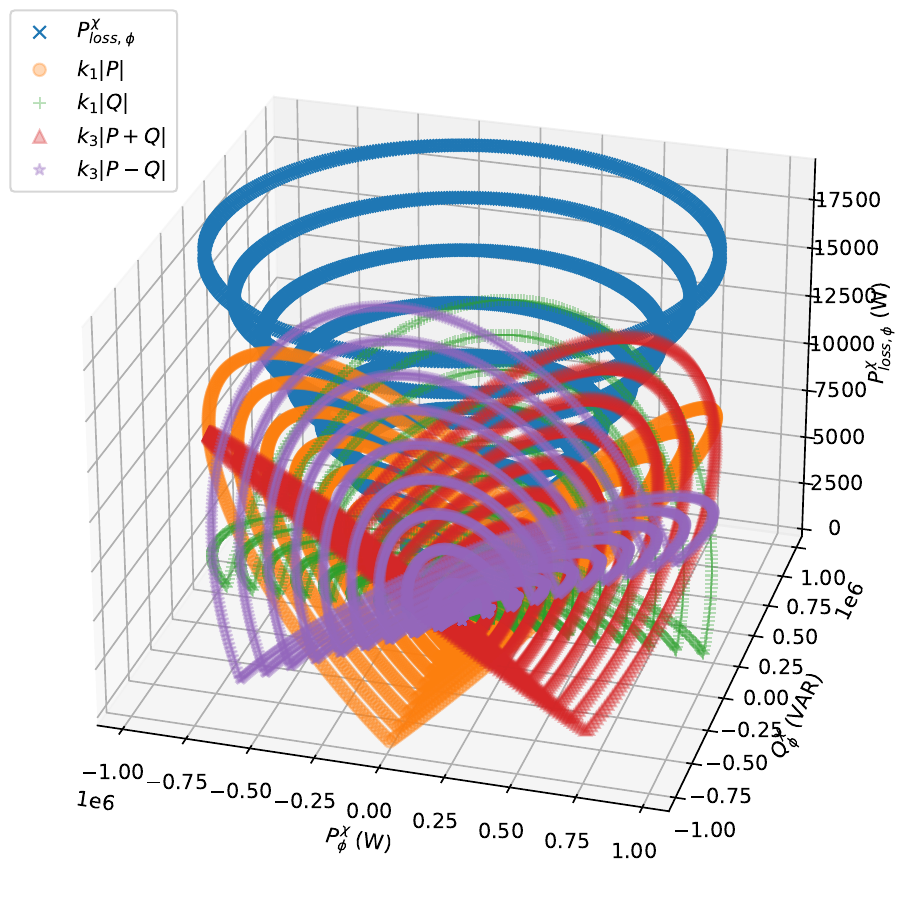}
  \caption{SOP loss surface with approximating planar surfaces}
  \label{fig:fig3}
\end{figure}
\noindent where $\Gamma$ is a device parameter. Some of the schemes proposed in the literature model the SOP as a lossless device  \cite{baiDistributedEnergyStorage2018} which is impractical whereas others propose an alternate expression for the power losses. The authors in \cite{louNewPhaseChanging2020} model PCSOP losses as a function of active power injections. This implies that the SOP functions as a lossless device when used as a condenser to provide reactive power support. The expression in (\ref{eq:3}) is therefore a more plausible representation of device losses over the entire operating range of the SOP. Considering the expression for power loss in (\ref{eq:3}) and the associated visual representation of the loss surface in figure \ref{fig:fig3}, we propose $f^{\chi}_{loss,\phi}$ as a functional approximation to $P^{\chi}_{loss,\phi}$.
\begin{equation}
\begin{aligned}    
\label{eq:4}
f^{\chi}_{loss,\phi} = &k^{\chi}_{\phi,1}|P^{\chi}_{\phi}|+k^{\chi}_{\phi,2}|Q^{\chi}_{\phi}|+k^{\chi}_{\phi,3}|P^{\chi}_{\phi}+Q^{\chi}_{\phi}|\\
&+k^{\chi}_{\phi,4}|P^{\chi}_{\phi}-Q^{\chi}_{\phi}|+k^{\chi}_{\phi,5}\\
&\approx P^{\chi}_{loss,\phi}
\end{aligned}
\end{equation}
The SOP loss function can be potentially expressed as a linear combination of terms involving active and reactive power injections. The reformulation proposed in (\ref{eq:4}) is motivated by the following:
\begin{itemize}
    \item The approximation given by $f^{\chi}_{loss,\phi}$ is mathematically convenient and geometrically interpretable. The proposed formulation expresses the SOP power losses as a weighted combination of planar surfaces depicted in figure \ref{fig:fig3}. The set of weight coefficients $[k^{\chi}_{\phi,1}...k^{\chi}_{\phi,5}]$ can be learned by sampling $P^{\chi}_{loss,\phi}$. 
    \item The expression in (\ref{eq:4}), which is still nonlinear in the context of mathematical programming since it relies on absolute values of component terms, is amenable to a well-known transformation that helps linearize $f^{\chi}_{loss,\phi}$.  
\end{itemize}
The approximation proposed in (\ref{eq:4}) and the equality constraints in (\ref{eq:1}) can be reformulated as a set of linear equality and inequality constraints \cite{bertsimasIntroductionLinearOptimization1997} by introducing auxiliary variables associated with the terms in (\ref{eq:4}). Consider the variable substitutions and bounds introduced in (\ref{eq:5}) and (\ref{eq:6}).
\begin{equation}
\begin{aligned}    
\label{eq:5}
|P^{\chi}_{\phi}| &= U^{\chi}_{\phi,P}\\
|Q^{\chi}_{\phi}| &= U^{\chi}_{\phi,Q}\\
|P^{\chi}_{\phi}+Q^{\chi}_{\phi}| &= U^{\chi}_{\phi,P+Q}\\
|P^{\chi}_{\phi}-Q^{\chi}_{\phi}| &= U^{\chi}_{\phi,P-Q}
\end{aligned}
\end{equation}
\begin{equation}
\begin{aligned}    
\label{eq:6}
-U^{\chi}_{\phi,P}\leq &P^{\chi}_{\phi} \leq U^{\chi}_{\phi,P}\\
-U^{\chi}_{\phi,Q}\leq &Q^{\chi}_{\phi}\leq U^{\chi}_{\phi,Q}\\
-U^{\chi}_{\phi,P+Q}\leq P^{\chi}_{\phi}&+Q^{\chi}_{\phi}\leq U^{\chi}_{\phi,P+Q}\\
-U^{\chi}_{\phi,P-Q}\leq P^{\chi}_{\phi}&-Q^{\chi}_{\phi}\leq U^{\chi}_{\phi,P-Q}
\end{aligned}
\end{equation}
where $U^{\chi}_{\phi,P}$, $U^{\chi}_{\phi,Q}$, $U^{\chi}_{\phi,P+Q}$, and $U^{\chi}_{\phi,P-Q}$ are the set of auxiliary variables. The function $f^{\chi}_{loss,\phi}$ can now be expressed using the continuous auxiliary variables.  
\begin{equation}
\begin{aligned}    
\label{eq:7}
f^{\chi}_{loss,\phi} = &k^{\chi}_{\phi,1}U^{\chi}_{\phi,P}+k^{\chi}_{\phi,2}U^{\chi}_{\phi,Q}+k^{\chi}_{\phi,3}U^{\chi}_{\phi,P+Q}\\
&+k^{\chi}_{\phi,4}U^{\chi}_{\phi,P-Q}+k^{\chi}_{\phi,5}\\
&\approx P^{\chi}_{loss,\phi}
\end{aligned}
\end{equation}
\begin{figure}[!t]
  \centering
  \includegraphics[scale=0.33]{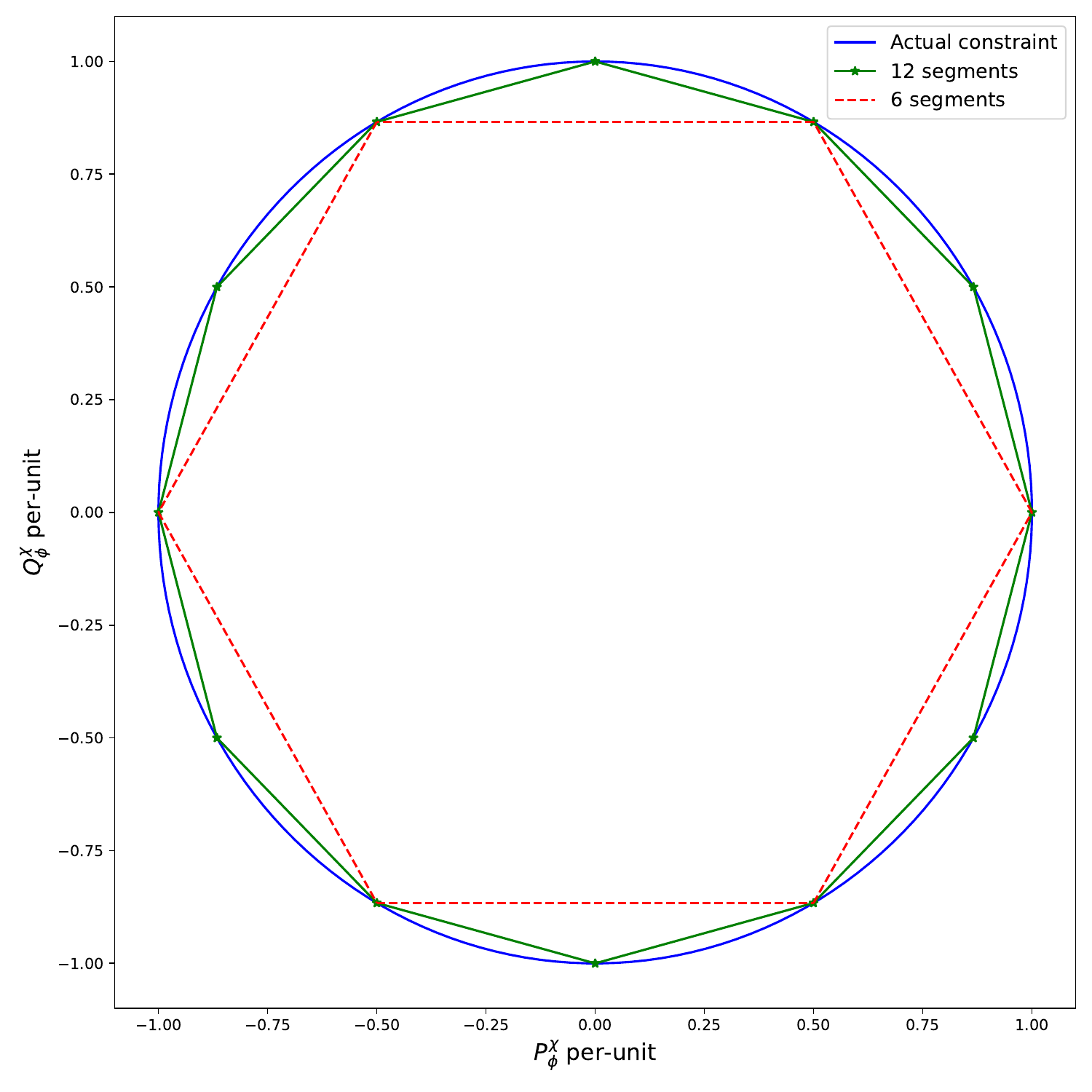}
  \caption{Capacity constraints approximated by line segments}
  \label{fig:fig4}
\end{figure}
\noindent Substituting the expressions in (\ref{eq:5})-(\ref{eq:7}) into (\ref{eq:1}) linearizes the active power balance constraints. The SOP capacity constraints given by (\ref{eq:2}) are also nonlinear and need to be linearized to accelerate program convergence. This can be done by approximating the circular device capacity constraints by a series of line segments that span the periphery of the device limits as shown in figure \ref{fig:fig4}. The capacity constraints can be approximated to an arbitrary accuracy by setting the desired number of line segments. The equations of the line segments are given by the expression in (\ref{eq:8}).
\begin{equation}
\begin{aligned}    
\label{eq:8}
Q^{\chi,i}_{\phi}-k^i_sP^{\chi,i}_{\phi}-&S^{\chi}_{\phi,max}[sin(\theta_j)-k^i_scos(\theta_j)]=0\\
&k^i_s = \frac{sin(\theta_{j+1})-sin(\theta_j)}{cos(\theta_{j+1})-cos(\theta_j)}
\end{aligned}
\end{equation}
where $Q^{\chi,i}_{\phi}$ and $P^{\chi,i}_{\phi}$ are the reactive and active power injections along the line segment $i$, and $\theta_j$ and $\theta_{j+1}$ are angles at the two extremes of line segment $i$ measured counter-clockwise from the active power axis. The set of expressions presented in (\ref{eq:5})-(\ref{eq:8}) linearize the operational dynamics of SOPs. 

The weight coefficients $[k^{\chi}_{\phi,1}...k^{\chi}_{\phi,5}]$ can be obtained using linear regression. The terms $|P^{\chi}_{\phi}|$, $|Q^{\chi}_{\phi}|$, $|P^{\chi}_{\phi}+Q^{\chi}_{\phi}|$ and $|P^{\chi}_{\phi}-Q^{\chi}_{\phi}|$ are treated as features. We uniformly sample the loss function surface to obtain a set of features and targets.
\begin{equation}
\begin{aligned}    
\label{eq:9}
([P^{\chi}_{\phi,i},Q^{\chi}_{\phi,i}],y_i)\sim P^{\chi}_{loss,\phi}
\end{aligned}
\end{equation}
where $y_i$ are the target (actual loss) values. Computing the parameters using standard gradient descent leads to $f^{\chi}_{loss,\phi}$ that closely tracks the actual SOP loss function. The update equation for the parameters $k^{\chi}_{\phi,i}$ is given by:
\begin{equation}
\label{eq:10}
    k^{\chi}_{\phi,i}:=k^{\chi}_{\phi,i}-\nabla_{k^{\chi}_{\phi,i}}(\frac{1}{N}\sum^{N}_{j=1}(y_j-f^{\chi}_{loss,\phi}(P^{\chi}_{\phi,j},Q^{\chi}_{\phi,j}))^2) 
\end{equation}
\noindent The following subsections succinctly describe the mathematical modeling of the case study applications with the proposed linearization procedure embedded in the optimal power flow (OPF) programs. The case studies were designed to assess the proposed framework for convergence and execution times. 

\subsection{Optimal Sizing of PCSOP and Minimizing Network Losses}
The first case study involves optimal sizing of the PCSOP for reducing network losses. The second case study is similar except that the per-phase capacity of the PCSOP is given as a parameter and the application involves extracting the set of optimal power injections for a given load scenario. The case studies were designed for the system depicted in figure \ref{fig:fig1}. The objective functions for the two case studies are given by (\ref{eq:11}) and (\ref{eq:12}), respectively.  
\begin{equation}
\label{eq:11}
    \underset{S^{\chi}_{\phi,max}}{min} \ \sum\limits_{l\in \Omega_{L},\phi \in \Omega_{\phi}}P_{loss,l}^{\phi}+\sum\limits_{\chi\in \Omega_{\chi},\phi \in \Omega_{\phi}}P^{\chi}_{loss,\phi}  
\end{equation}
\begin{equation}
\label{eq:12}
   \underset{P^{\chi}_{\phi},Q^{\chi}_{\phi}}{min} \ \sum\limits_{l\in \Omega_{L},\phi \in \Omega_{\phi}}P_{loss,l}^{\phi}+\sum\limits_{\chi\in \Omega_{\chi},\phi \in \Omega_{\phi}}P^{\chi}_{loss,\phi}   
\end{equation}
where $P_{loss,l}^{\phi}$ is the power-loss in phase $\phi$ of line segment $l$, $\Omega_{L}$ is the set of all line segments in the distribution grid, $\Omega_{\chi}$ is the set of all PCSOP bus terminals, and $\Omega_{\phi}$ is the set of phases. The power flow equations for the three-phase unbalanced network account for the mutual coupling that may exist between phases. The complex-valued phase power flowing from bus $k$ to $m$ is given by $S^{km}_{\phi}$.
\begin{equation}
\label{eq:13}
\begin{aligned}
    S^{km}_{\phi} = V^{k}_{\phi}&\sum_{i \in (a,b,c)}(y^{km}_{\phi i})^*(V^{k}_{i})^*-(y^{km}_{\phi i})^*(V^{m}_{i})^* \\ 
    &\forall\phi \in (a,b,c),\ (k,m) \in \Omega_{B}  
\end{aligned}
\end{equation} 
where $V^{k}_{\phi}$ is the bus phase voltage, $y^{km}_{\phi i}$ is a component of the bus admittance matrix, and $\Omega_{B}$ is the set of all network buses. Loads in the case studies are modeled as polynomial (ZIP) loads where the active ($p^k_{load,\phi}$) and reactive ($q^k_{load,\phi}$) powers consumed are functions of the nominal load power, the load polynomial coefficients, the nominal bus voltage, and the actual bus voltage. Bus voltage constraints in (\ref{eq:14}) ensure the system operates within the acceptable voltage range.
\begin{equation}
\label{eq:14}
0.95\leq |V^{k}_{\phi}|\leq 1.05, \ \forall \phi \in (a,b,c), \ k \in \Omega_{B}
\end{equation}
The OPF programs involve minimizing the objective in (\ref{eq:11}) or (\ref{eq:12}) subject to the constraints established in (\ref{eq:5})-(\ref{eq:8}) and (\ref{eq:13})-(\ref{eq:14}).        

\subsection{Case Study on Reinforcement Learning}
This case study was designed to assess the performance of the proposed SOP representations when embedded into an online process simulator. Reinforcement learning (RL) applications require environment representations that can not only simulate state transitions and reward emissions with reasonable accuracy but do so within the allotted time budget. In this application, an agent was trained to optimally dispatch the PCSOP using the Deep Deterministic Policy Gradient (DDPG) \cite{lillicrapContinuousControlDeep2019} algorithm and minimize network losses. DDPG is an off-policy RL method that implements a continuous deterministic policy and trains an agent by co-optimizing the actor (policy) and critic neural networks. The critic is trained to estimate the discounted return (Q-value) for a given state-action pair and the policy network is trained to output action (active and reactive power injection) vectors that maximize the Q-value. The policy objective function is given by:
\begin{equation}
    \label{eq:15}
    \psi^* = arg \ \underset{\psi}{max} \underset{s\sim D}{\E}[Q_{\theta}(s,\mu_{\psi}(s))]
\end{equation}
where $\mu_{\psi}$ is the policy network parametrized by parameter set $\psi$, $s$ is the system state, $Q_{\theta}$ is the critic network parametrized by $\theta$, and $D$ is the set of experiences. The input to the policy network (state) is the net power injection at each bus of the distribution network in figure \ref{fig:fig1}. The reward function used to steer the agent is analogous to system power efficiency where injections from the PCSOP that lower network losses are rewarded on a continuous scale. The core of the environment simulation engine is the nonlinear AC load flow routine implemented with the linearized representation of SOP dynamics. Actions sampled from the policy network are passed on to the simulator as parameters. The process simulator executes the load flow routine using the system state and action vector and returns the associated reward to update the policy parameters.     

\section{Performance Evaluation}
The section presents the results of the calculation of the weighting parameters $[k^{\chi}_{\phi,1}...k^{\chi}_{\phi,5}]$. The weighting coefficients obtained were then used to assess the performance of the proposed framework in the case study applications. 

\subsection{Parameter Calculation for $f^{\chi}_{loss,\phi}$}
A dataset of samples was first obtained by uniformly sampling the loss function surface shown in figure \ref{fig:fig3}. The parameters were calculated using the standard linear regression algorithm with the update equation in (\ref{eq:10}). Table \ref{tab:1} lists the training parameters and values of the coefficients obtained for one phase of an SOP terminal. The device capacities $S^{\chi}_{\phi,max}$ and loss parameter $\Gamma$ are conventionally presumed to be identical for all terminals and phases of an SOP. It is worth noting that the proposed approximation is scale-invariant. A change in device capacity does not require recalculation of the coefficients in (\ref{eq:4}).
\begin{table}[!t]
\begin{center}
\caption{Training Parameters and Coefficient Values.}
\label{tab:1}
\begin{tabular}{| c | c |}
\hline
Parameter & Value \\
\hline
Per-phase Capacity of SOP& 1 MVA\\
\hline
$\Gamma$& 0.02\\
\hline
Dataset size& 250000\\
\hline
$k_1$& 0.0078\\
\hline
$k_2$& 0.0078\\
\hline
$k_3$& 0.0055\\
\hline
$k_4$& 0.0055\\
\hline
$k_5$& 0.0177\\
\hline 
\end{tabular}
\end{center}
\end{table}
\begin{figure}[!t]
  \centering
  \includegraphics[scale=0.5]{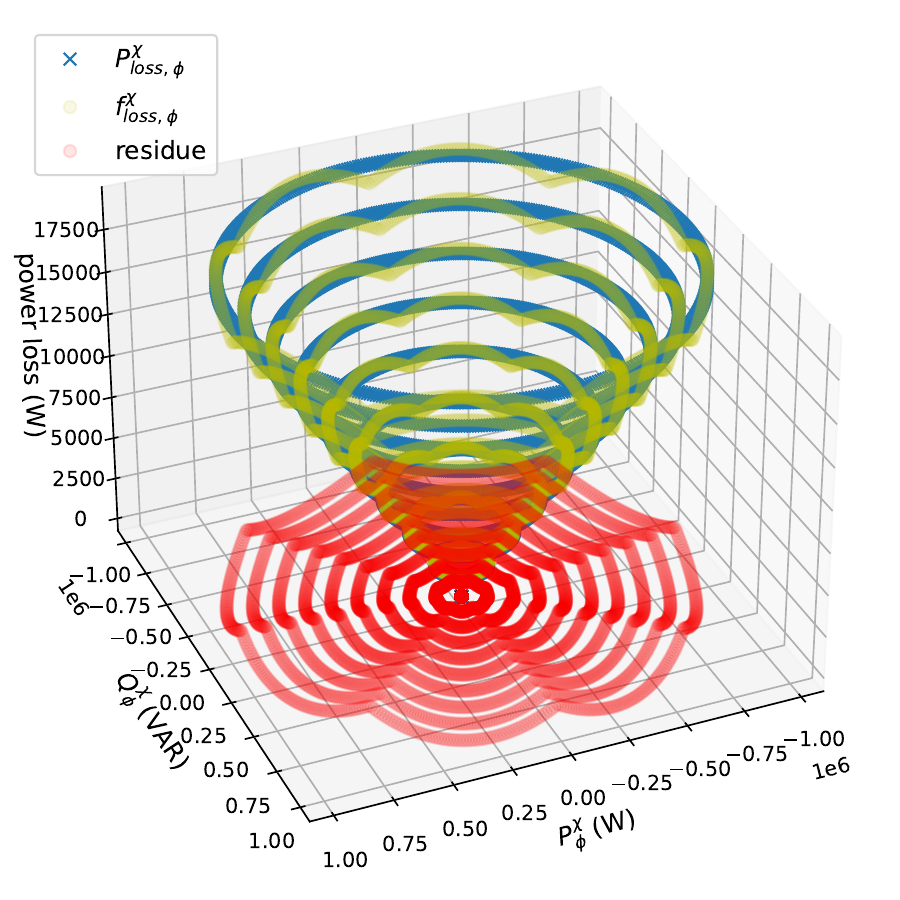}
  \caption{Actual and approximated loss surfaces}
  \label{fig:fig5}
\end{figure}
The plot in figure \ref{fig:fig5} shows the result of approximating the nonlinear loss function using the parameters calculated in Table \ref{tab:1}. The linearized loss surface closely matches the actual losses. In fact, the maximum approximation error is less than 6\% across the entire operating range of the SOP with a capacity of 1 MVA.

\subsection{Results of SOP Optimal Sizing and Operation Optimization}
Optimal sizing of the SOP is important for planning, expansion, and commissioning of distribution networks. The case studies were designed for the unbalanced test-feeder shown in figure \ref{fig:fig1}. The OPF programs and load flow routines were implemented using the open-source PYOMO optimization modeling language and solved using the efficient IPOPT solver. The optimal sizing application calculates the smallest feasible per-phase capacity of the PCSOP under maximum loading conditions with the DGs turned off. The output of this case study was then used to execute the OPF program and load flow routines of the other two applications. Table \ref{tab:2} lists the system parameters used to run the optimal sizing and operational optimization applications. The system parameters for the RL application are identical. The results of running the optimal sizing application are reported in Table \ref{tab:3}.   
\begin{table}[!t]
\begin{center}
\caption{System Parameters.}
\label{tab:2}
\begin{tabular}{| c | c |}
\hline
Parameter & Value \\
\hline
System voltage& 12.66 kV\\
\hline
Nominal system load& 11.6+j6.96 MVA\\
\hline
DG total peak generation& 2.75+j0.9 MVA\\
\hline 
\end{tabular}
\end{center}
\end{table}
\begin{table}[!t]
\begin{center}
\caption{Output of SOP Sizing Application.}
\label{tab:3}
\begin{tabular}{| c | c | c |}
\hline
Variable&Value with&Value using\\
&proposed method&conventional method\\
\hline
Calculated SOP size & 0.967 MVA & \multirow{5}{*}{Failed to converge}\\
\cline{1-2}
Network power loss& 3.47\% &\\
using $P^{\chi}_{loss,\phi}$ & & \\
\cline{1-2}
Network power loss& 3.44\% & \\
using $f^{\chi}_{loss,\phi}$ & & \\
\cline{1-2}
Average solution time& 6.3 sec & \\
\cline{1-2}
Bus voltage violations & None & \\
\hline 
\end{tabular}
\end{center}
\end{table}
\begin{table}[!t]
\begin{center}
\caption{Output of Operational Optimization Application.}
\label{tab:4}
\begin{tabular}{| c | c | c |}
\hline
Loading condition & System losses using & System losses using\\
& $f^{\chi}_{loss,\phi}$ & $P^{\chi}_{loss,\phi}$ \\ 
\hline
Minimum load & 1.73\% & 1.75\%\\
\hline
Maximum load& 2.4\% & 2.43\%\\
\hline
Minimum loss loading& 1.63\% & 1.66\%\\
\hline
Maximum loss loading& 2.42\% & 2.45\%\\
\hline 
\end{tabular}
\end{center}
\end{table}
Combining the nonlinear power flow formulation in AC OPF programs with the standard representation of SOP dynamics induces a solution search space that is extremely challenging to navigate. In practice, the optimization programs invariably fail to converge. In contrast, the optimal sizing application designed using the proposed SOP linearization procedure efficiently converged to an optimal solution within a few seconds. The total system power loss calculated using the actual loss function in (\ref{eq:3}) and the proposed approximation were in close agreement as shown in Table \ref{tab:3}.

The second case study application involved mitigating the operational power losses in the unbalanced network for a given system load. A dataset of 500 system loads was generated by independently varying the per-phase bus loads and the DG power injections across the distribution network. Given the system load and the PCSOP capacity, the optimization program resolved the set of optimal PCSOP power injections that minimized network losses. The results of the case study are reported in Table \ref{tab:4}. Operational optimization application executed using the conventional OPF program failed to converge and the results are omitted. Once again, the system losses calculated using the proposed method closely track the actual losses. On average, the case study required a solution time of 0.62 seconds per iteration.    

\subsection{Results of Reinforcement Learning Application}
\begin{figure}[!t]
  \centering
  \includegraphics[scale=0.5]{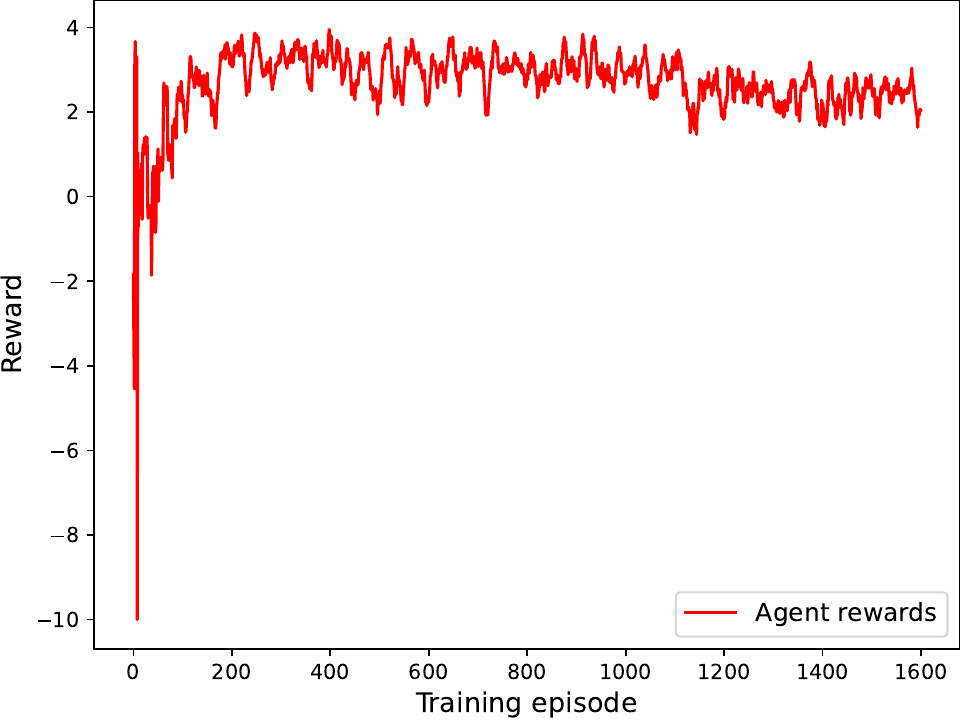}
  \caption{Training of agent using DDPG}
  \label{fig:fig6}
\end{figure}

\begin{figure}[!t]
  \centering
  \includegraphics[scale=0.5]{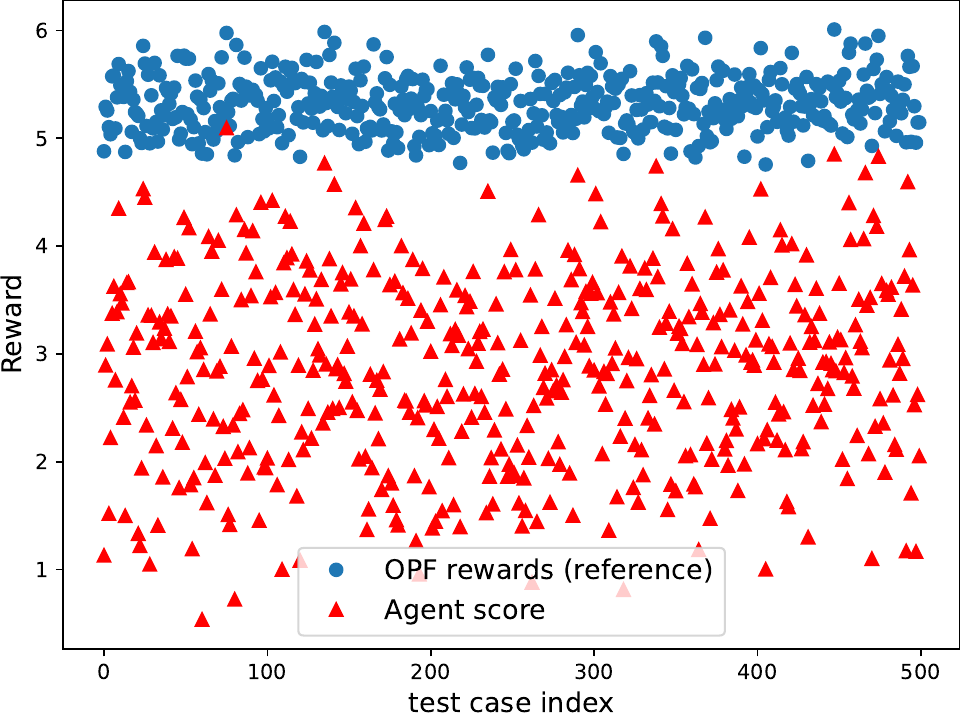}
  \caption{Comparison of agent's test performance}
  \label{fig:fig7}
\end{figure}

The environment simulator designed using the proposed framework was used to train an agent using the DDPG algorithm. DDPG is a model-free RL method where an agent learns to approximate the process dynamics by repeated interactions with the environment. The application was implemented using the PyTorch deep learning library. Table \ref{tab:5} details the framework parameters used to define and train the agent. The agent was trained using the training set and its performance was analyzed on a separate test set. The parameter $\sigma$ is the standard deviation value of the Gaussian distribution used to explore the action space.    
\begin{table}
\begin{center}
\caption{Agent Training Parameters.}
\label{tab:5}
\begin{tabular}{| c | c |}
\hline
Parameter & Value\\
\hline
Hidden layers& 2\\
(actor, critic)& \\
\hline
Neurons per layer&64\\ 
\hline
learning rate& 4e-4\\
\hline
Mini-batch size& 128\\
\hline
Exploration Noise ($\sigma$) & 0.25\\
\hline
Noise floor ($\sigma$)& 0.1\\
\hline
Optimizer& Adaptive Moment Estimation\\
\hline
Training set size & 4500\\
\hline
Test set size & 500\\
\hline
Training episodes & 1600\\
\hline 
\end{tabular}
\end{center}
\end{table}
Figure \ref{fig:fig6} depicts the evolution of rewards over the course of training episodes and figure \ref{fig:fig7} compares the performance of the agent on instances from the test set with the OPF-based reference scores. The scalar reward is analogous to the overall system power efficiency. A higher reward signifies an improvement in the agent's ability to optimally dispatch the PCSOP. The case study was designed to assess the performance of the proposed SOP representation when embedded into an online process simulator.         
Benchmarking the performance of the agent using the conventional OPF program led to repeated stalling of the training process due to the process simulator failing to converge. The process had to be eventually abandoned and was re-executed with an OPF program embedded with the proposed SOP representation. Figure \ref{fig:fig7} shows the trained agent lagging behind the reference scores on the test set. This was mainly due to the complexity of the action space associated with the operation of the SOP and will be addressed in a later investigation.    

\section{Conclusion}
This paper proposed a novel data-driven, regression-based reformulation of the SOP power-loss functions. The linearization technique transforms the nonlinear power-loss expressions into a linear combination of the active and reactive power injection variables. The proposed procedure, combined with the well-known linear segmentation of quadratic device capacity constraints, linearizes the operational dynamics of SOPs without a significant loss in accuracy. The proposed representation was tested in three case studies that are representative of planning, operations optimization, and data-driven control applications.    


%
\bibliographystyle{IEEEtran}
\bibliography{ref.bib}



\end{document}